\documentclass[screen,sigconf]{acmart}
\AtBeginDocument{%
  \providecommand\BibTeX{{%
    \normalfont B\kern-0.5em{\scshape i\kern-0.25em b}\kern-0.8em\TeX}}}

\acmYear{2021}\copyrightyear{2021}
\setcopyright{rightsretained}
\acmConference[ACM CHIL '21]{ACM Conference on Health, Inference, and Learning}{April 8--10, 2021}{Virtual Event, USA}
\acmBooktitle{ACM Conference on Health, Inference, and Learning (ACM CHIL '21), April 8--10, 2021, Virtual Event, USA}
\acmPrice{}
\acmDOI{10.1145/3450439.3451876}
\acmISBN{978-1-4503-8359-2/21/04}

\begin{document}

\title[CheXternal: Generalization of Deep Learning Models for Chest X-ray Interpretation]{CheXternal: Generalization of Deep Learning Models for Chest X-ray Interpretation to Photos of Chest X-rays and External Clinical Settings}
\author{Pranav Rajpurkar}
\authornote{Authors contributed equally to this research.}
\email{pranavsr@cs.stanford.edu}
\affiliation{
    \institution{Stanford University}
    % \streetaddress{353 Jane Stanford Way}
    % \city{Stanford}
    % \state{California}
    \country{USA}
    % \postcode{94305}
}
\author{Anirudh Joshi}
\authornotemark[1]
\email{anirudhjoshi@cs.stanford.edu}
\affiliation{
    \institution{Stanford University}
    % \streetaddress{353 Jane Stanford Way}
    % \city{Stanford}
    % \state{California}
    \country{USA}
    % \postcode{94305}
}
\author{Anuj Pareek}
\authornotemark[1]
\email{anujpare@cs.stanford.edu}
\affiliation{
    \institution{Stanford University}
    % \streetaddress{353 Jane Stanford Way}
    % \city{Stanford}
    % \state{California}
    \country{USA}
    % \postcode{94305}
}
\author{Andrew Y. Ng}
\email{ang@cs.stanford.edu}
\affiliation{
    \institution{Stanford University}
    % \streetaddress{353 Jane Stanford Way}
    % \city{Stanford}
    % \state{California}
    \country{USA}
    % \postcode{94305}
}
\author{Matthew P. Lungren}
\email{mlungren@stanford.edu}
\affiliation{
    \institution{Stanford University}
    % \streetaddress{353 Jane Stanford Way}
    % \city{Stanford}
    % \state{California}
    \country{USA}
    % \postcode{94305}
}
% \renewcommand{\shortauthors}{Rajpurkar, Joshi, Pareek et al.}


\begin{abstract}
Recent advances in training deep learning models have demonstrated the potential to provide accurate chest X-ray interpretation and increase access to radiology expertise. However, poor generalization due to data distribution shifts in clinical settings is a key barrier to implementation. In this study, we measured the diagnostic performance for 8 different chest X-ray models when applied to (1) smartphone photos of chest X-rays and (2) external datasets without any finetuning. All models were developed by different groups and submitted to the CheXpert challenge, and re-applied to test datasets without further tuning. We found that (1) on photos of chest X-rays, all 8 models experienced a statistically significant drop in task performance, but only 3 performed significantly worse than radiologists on average, and (2) on the external set, none of the models performed statistically significantly worse than radiologists, and five models performed statistically significantly better than radiologists. Our results demonstrate that some chest X-ray models, under clinically relevant distribution shifts, were comparable to radiologists while other models were not. Future work should investigate aspects of model training procedures and dataset collection that influence generalization in the presence of data distribution shifts.  
\end{abstract}




\begin{teaserfigure}
\centering
  \includegraphics[width=0.8\textwidth]{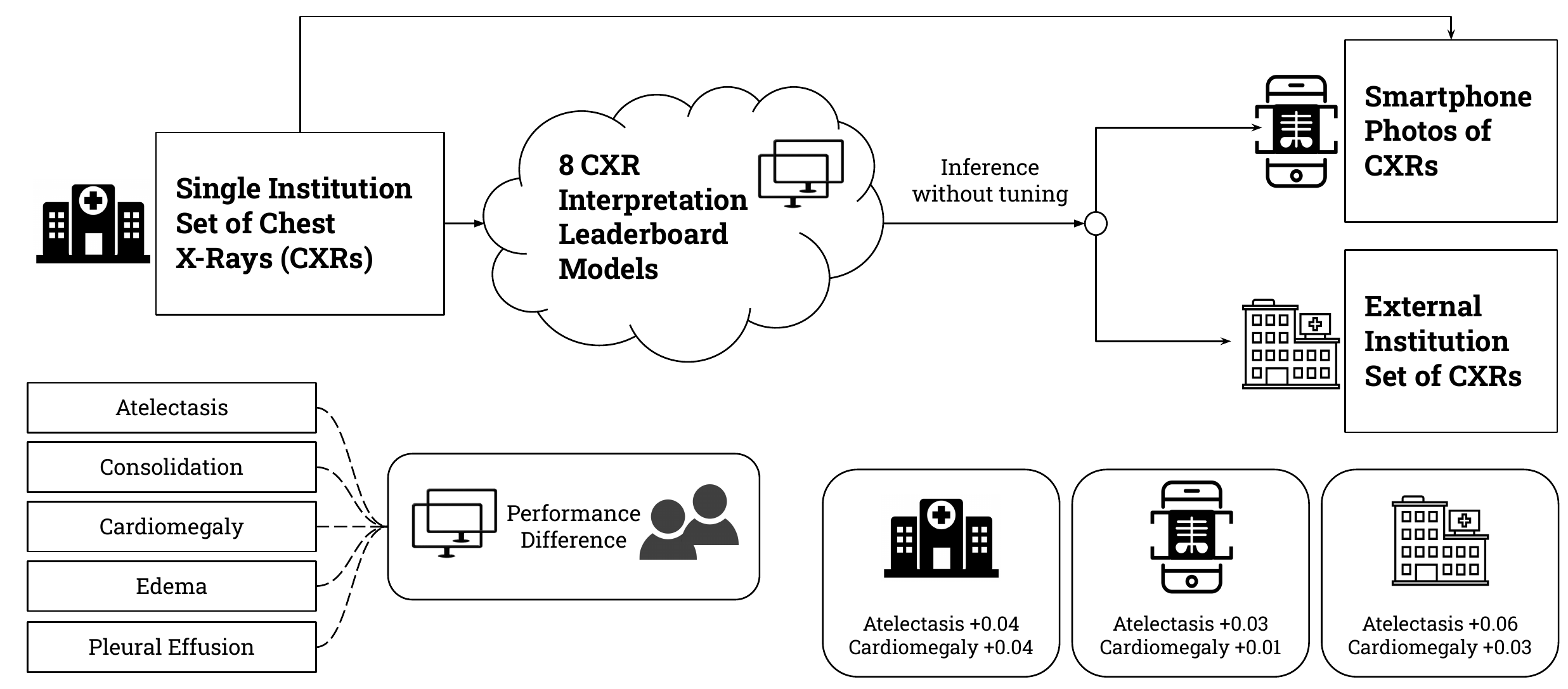}
  \caption{We measured the diagnostic performance for 8 different chest X-ray models when applied to (1) smartphone photos of chest X-rays and (2) external datasets without any finetuning. All models were developed by different groups and submitted to the CheXpert challenge, and re-applied to test datasets without further tuning.}
  \label{fig:teaser}
\end{teaserfigure}

\maketitle

\section{Introduction}
\label{sec:intro}

Chest X-rays are the most common imaging examination in the world, critical for diagnosis and management of many diseases. With over 2 billion chest X-rays performed globally annually, many clinics in both developing and developed countries lack sufficient trained radiologists to perform timely X-ray interpretation. Automating cognitive tasks in medical imaging interpretation with deep learning models could improve access, efficiency, and augment existing workflows \citep{rajpurkar_deep_2018, nam_development_2018, singh_deep_2018, qin_computer-aided_2018}. However, poor generalization due to data distribution shifts in clinical settings is a key barrier to implementation.

First, a major obstacle to clinical adoption of such technologies is in model deployment, an effort often frustrated by vast heterogeneity of clinical workflows across the world \citep{kelly_key_2019}. Chest X-ray models are developed and validated using digital X-rays with many deployment solutions relying on heavily integrated yet often disparate infrastructures \citep{qin_using_2019, lakhani_deep_2017, kallianos_how_2019, kashyap_artificial_2019, shih_augmenting_2019, andronikou2011paediatric,schwartz_accuracy_2014}. 
One appealing solution to scaled deployment across disparate clinical frameworks is to leverage the ubiquity of smartphones.  Interpretation of medical imaging via cell phone photography is an existing ``store-and-forward telemedicine'' approach in which one or more photos of medical imaging are captured and sent as email attachments or instant messages by practitioners to obtain second opinions from specialists in routine clinical care \citep{goost2012image, vassallo1998first}. 
Smartphone photographs have been shown to be of sufficient diagnostic quality to allow for medical interpretation, thus leveraging deep learning models in automated interpretation of photos of medical imaging examinations may serve as an infrastructure agnostic approach to deployment, particularly in resource limited settings. However, significant technical barriers exist in automated interpretation of photos of chest X-rays. Photographs of X-rays introduce visual artifacts which are not commonly found in digital X-rays, such as altered viewing angles, variable lighting conditions, glare, moiré, rotations, translations, and blur \cite{phillips2020chexphoto}. These artifacts have been shown to reduce algorithm performance when input images are perceived through a camera \citep{kurakin_adversarial_2016}. The extent to which such artifacts reduces the performance of chest X-ray models has not been well investigated.

A second major obstacle to clinical adoption of chest X-ray models is that clinical deployment requires models trained on data from one institution to generalize to data from another institution \citep{kelly_key_2019, chen_deep_2019}. Early work has shown that chest X-ray models may not generalize well when externally validated on data from a different institution and are possibly vulnerable to distribution shift stemming from change in patient population or rely on non-medically relevant cues between institutions \citep{zech_variable_2018}. However, the difference in diagnostic performance of more recent chest X-ray models to external datasets has not been investigated. 

We measured the diagnostic performance for 8 different chest X-ray models when applied to (1) photos of chest X-rays, and (2) chest X-rays obtained at a different institution. Specifically, we applied these models to a dataset of smartphone photos of 668 X-rays from 500 patients, and a set of 420 frontal chest X-rays from the ChestXray-14 dataset collected at the National Institutes of Health Clinical Center \citep{wang_chestX-ray8_2017}. All models were developed by different groups and submitted to the CheXpert challenge, a large public competition for digital chest X-ray analysis \citep{irvin_chexpert_2019}. Models were evaluated on their diagnostic performance in binary classification, as measured by Matthew’s Correlation Coefficient (MCC) \citep{chicco2020advantages}, on the following pathologies selected in \citet{irvin_chexpert_2019}: atelectasis, cardiomegaly, consolidation, edema, and pleural effusion \citep{irvin_chexpert_2019}.

We found that:

\begin{enumerate}
    \item In comparison of model performance on digital chest X-rays to photos, all 8 models experienced a statistically significant drop in task performance on photos with an average drop of 0.036 MCC. In comparison of performance of models on photos compared to radiologist performance, three out of eight models performed significantly worse than radiologists on average, and the other five had no significant difference.
    \item On the external set (NIH), none of the models performed statistically significantly worse than radiologists. On average over the pathologies, five models performed significantly better than radiologists. On specific pathologies (consolidation, cardiomegaly, edema, and atelectasis), there were some models that achieved significantly better performance than radiologists.
\end{enumerate}

Our systematic examination of the generalization capabilities of existing models can be extended to other tasks in medical AI, and provide a framework for tracking technical readiness towards clinical translation.

\begin{figure}[t]
\centering
  \includegraphics[width=0.8\columnwidth]{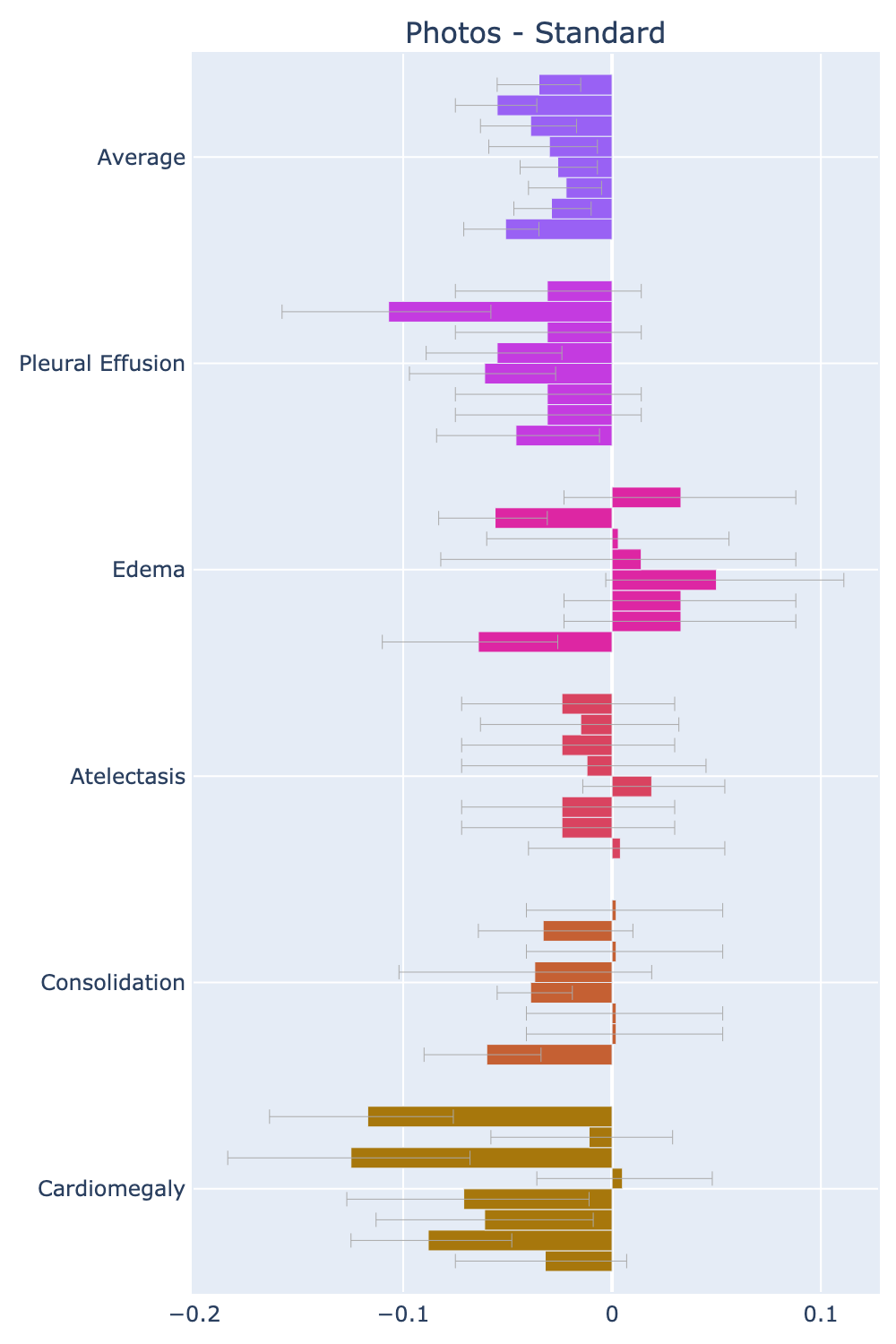}
  \caption{MCC differences of 8 chest X-ray models on different pathologies between photos of the X-rays and the original X-rays with 95\% confidence intervals.}
\label{fig:standard}
\end{figure}

\begin{figure}[t]
\centering
  \includegraphics[width=0.8\columnwidth]{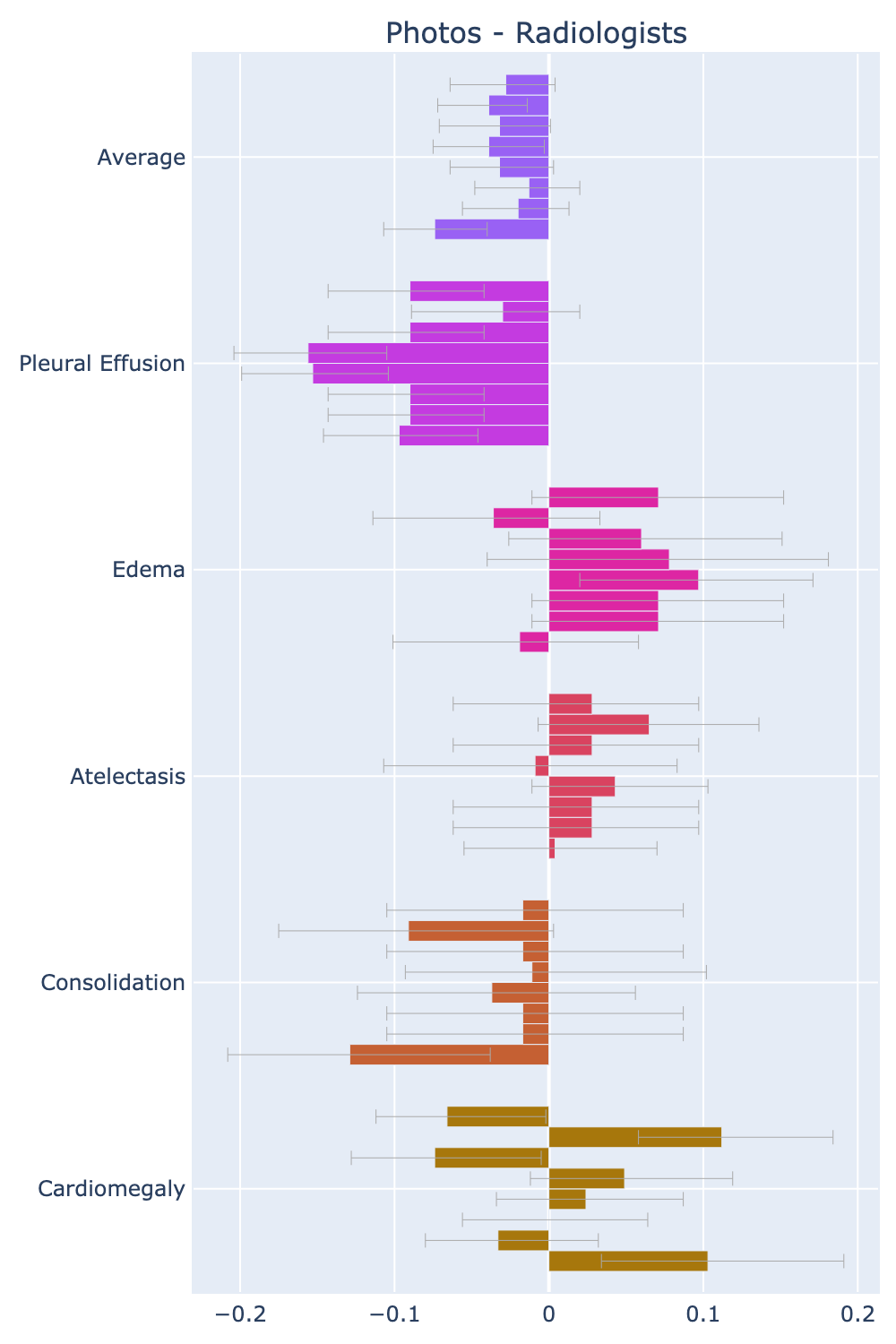}
  \caption{MCC differences of the same models on photos of chest X-rays compared to radiologist performance with 95\% confidence intervals. }
  \label{fig:photos}
\end{figure}

\begin{figure}[h]
\centering
  \includegraphics[width=\columnwidth]{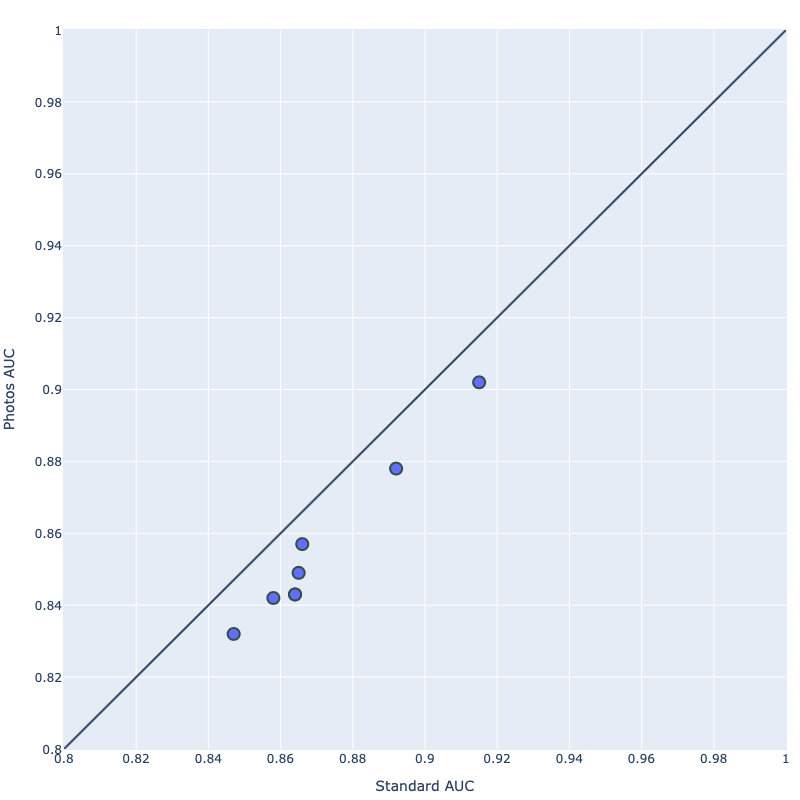}
  \caption{Comparison of the average AUC of 8 individual models on photos of chest X-rays compared to on standard images}
\label{fig:auc_comparison}
\end{figure}

\begin{table*}[h]
    \resizebox{\textwidth}{!}{  
\begin{tabular}{ll|r|rrrrr}
\toprule
Metric & Comparison        & Average             & Pleural   Effusion  & Edema                 & Atelectasis          & Consolidation        & Cardiomegaly        \\
\midrule
AUC    & Photos            & 0.856 (0.840,0.869) & 0.950 (0.932,0.968) & 0.917 (0.884,0.943)   & 0.882 (0.856,0.912)  & 0.914 (0.865,0.946)  & 0.921 (0.900,0.940) \\
    & Standard          & 0.871 (0.855,0.883) & 0.960 (0.944,0.975) & 0.926 (0.892,0.950)   & 0.885 (0.858,0.910)  & 0.918 (0.879,0.948)  & 0.934 (0.914,0.951) \\
    & Standard - Photos & 0.016 (0.012,0.019) & 0.011 (0.004,0.019) & 0.009 (0.001,0.018)   & 0.003 (-0.006,0.013) & 0.005 (-0.009,0.016) & 0.013 (0.006,0.023) \\
\midrule
MCC    & Photos            & 0.534 (0.507,0.559) & 0.571 (0.526,0.631) & 0.556 (0.481,0.639)   & 0.574 (0.505,0.634)  & 0.316 (0.246,0.386)  & 0.580 (0.522,0.630) \\
    & Standard          & 0.570 (0.543,0.599) & 0.621 (0.575,0.670) & 0.550 (0.474,0.637)   & 0.587 (0.529,0.640)  & 0.336 (0.264,0.418)  & 0.643 (0.584,0.695) \\
    & Standard - Photos & 0.036 (0.024,0.048) & 0.049 (0.020,0.070) & -0.006 (-0.039,0.033) & 0.012 (-0.016,0.041) & 0.020 (-0.011,0.047) & 0.063 (0.036,0.084)
    \\
\bottomrule
\end{tabular}
 }
  \caption{AUC and MCC performance of models and radiologists on the standard X-rays and the photos of chest X-rays, with 95\% confidence intervals.}
\label{tab:tab1}
\end{table*}

\begin{table*}[h]
    \resizebox{\textwidth}{!}{  
\begin{tabular}{lrrrrrr}
\toprule
Comparison              & Average             & Pleural   Effusion  & Edema                 & Atelectasis           & Consolidation        & Cardiomegaly          \\
\midrule
Photos                  & 0.534 (0.507,0.559) & 0.571 (0.526,0.631) & 0.556 (0.481,0.639)   & 0.574 (0.505,0.634)   & 0.316 (0.246,0.386)  & 0.580 (0.522,0.630)   \\
Radiologists            & 0.568 (0.542,0.597) & 0.671 (0.618,0.727) & 0.507 (0.431,0.570)   & 0.548 (0.496,0.606)   & 0.359 (0.262,0.444)  & 0.566 (0.511,0.620)   \\
\midrule
Radiologists -   Photos & 0.035 (0.009,0.065) & 0.099 (0.056,0.145) & -0.049 (-0.136,0.029) & -0.027 (-0.086,0.050) & 0.042 (-0.056,0.124) & -0.014 (-0.069,0.029) \\
\bottomrule
\end{tabular}
}
  \caption{MCC performance of models on the photos of chest X-rays, radiologist performance, and their difference, with 95\% confidence intervals.}
\label{tab:tab2}
\end{table*}

\section{Methods} 
\subsection{Photos of Chest X-rays}
We collected a test set of photos of chest x-rays, described in \citet{phillips2020chexphoto}. In this set, chest X-rays from each CheXpert test study were displayed on a non-diagnostic computer monitor. Chest X-rays were displayed in full screen on a computer monitor with $1920 \times 1080$ screen resolution and a black background. A physician was instructed to capture the photos, keeping the mobile camera stable and center the lung fields in the camera view. A time-restriction of 5 seconds per image was imposed to simulate a busy healthcare environment. Subsequent inspection of photos showed that they were taken with slightly varying angles; some photos included artefacts such as Moiré patterns and subtle screen glare. Photos were labeled using the ground truth for the corresponding digital X-ray image. The reference standard on this set was determined using a majority vote of 5 board-certified radiologists. Three separate board-certified radiologists were used for the comparison against the models and all radiologists used the original chest X-ray images for making their diagnoses, rather than the photos.

\subsection{Running Models on New Test Sets}
CheXpert used a hidden test set for official evaluation of models. Teams submitted their executable code, which was then run on a test set that was not publicly readable to preserve the integrity of the test results. We made use of the CodaLab platform to re-run these chest X-ray models by substituting the hidden CheXpert test set with the datasets used in this study. 

\subsection{Evaluation Metrics}
Our primary evaluation metric was Matthew’s Correlation Coefficient (MCC), a statistical rate which produces a high score only if the prediction obtained good results in all of the four confusion matrix categories (true positives, false negatives, true negatives, and false positives); MCC is proportionally both to the size of positive elements and the size of negative elements in the dataset \citep{chicco2020advantages}.

We reported the average MCC of 8 models for five pathologies, namely atelectasis, cardiomegaly, consolidation, edema, and pleural effusion. Additionally, in experiments comparing the models on standard chest X-rays to photos of chest X-rays, we reported the AUC and MCC of the models. In experiments comparing models to board-certified radiologists, we reported the difference in MCC for each of the five pathologies.

\section{Results}
\subsection{Model Performance on Photos of Chest X-rays vs Original X-rays.}

\subsubsection{Performance Drop In Application To Photos}
In comparison of model performance on digital chest X-rays to photos, all eight models experienced a statistically significant drop in task performance on photos with an average drop of 0.036 MCC (95\% CI 0.024, 0.048) (See Figure \ref{fig:standard}, Table \ref{tab:tab1}). All models had a statistically significant drop on at least one of the pathologies between native digital image to photos. One model had a statistically significant drop in performance on three pathologies: pleural effusion, edema, and consolidation. Two models had a significant drop on two pathologies: one on pleural effusion and edema, and the other on pleural effusion and cardiomegaly. The cardiomegaly and pleural effusion tasks led to decreased performance in five and four models respectively.

\subsubsection{Performance on Photos In Comparison to Radiologist Performance on Standard Images}
In comparison of performance of models on photos compared to radiologist performance, three out of eight models performed significantly worse than radiologists on average, and the other five had no significant difference (see Figure \ref{fig:photos}). On specific pathologies, there were some models that had a significantly higher performance than radiologists: two models on cardiomegaly, and one model on edema. Conversely, there were some models that had a significantly lower performance than radiologists: two models on cardiomegaly, and one model on consolidation.  The pathology with the greatest number of models that had a significantly lower performance than radiologists was pleural effusion (seven models).

\subsubsection{Performance drop in context of radiologist performance}
Our results demonstrated that while most models experienced a significant drop in performance when applied to photos of chest X-rays compared to the native digital image, their performance was nonetheless largely equivalent to radiologist performance. We found that although there were thirteen times that models had a statistically significant drop in performance on photos on the different pathologies, the models had significantly lower performance than radiologists only 6 of those 13 times. Comparison to radiologist performance provides context in regard to clinical applicability: several models remained comparable to radiologist performance standard despite decreased performance on photos. Further investigation could be directed towards understanding how different model training procedures may affect model generalization to photos of chest X-rays, and understanding etiologies behind trends for changes in performance for specific pathologies or specific artifacts.

\begin{figure}[t]
\centering
  \includegraphics[width=0.8\columnwidth]{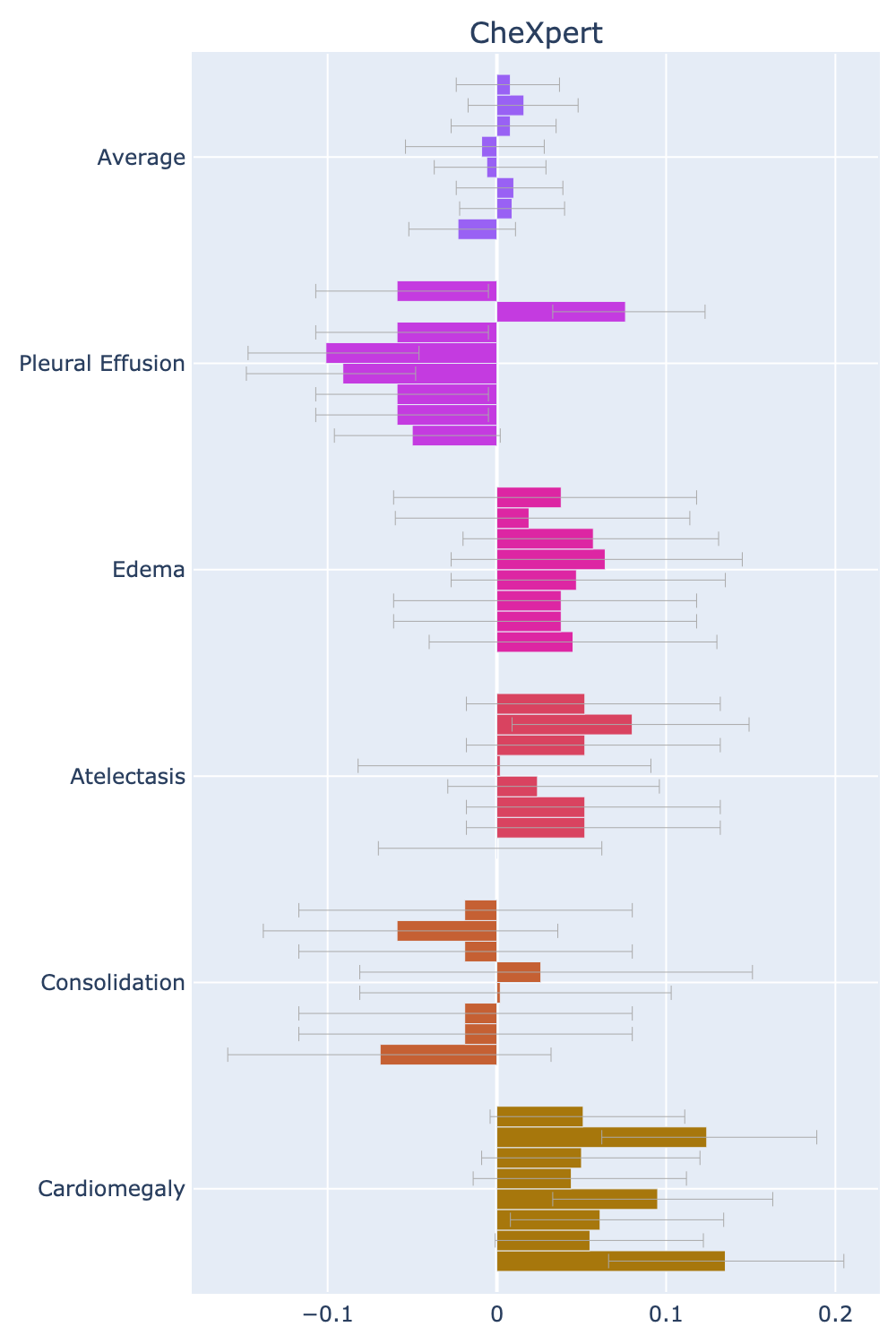}
  \caption{MCC differences in performance of models on the CheXpert test set, with 95\% confidence intervals (higher than 0 is in favor of the models being better).}
\label{fig:chexpert}
\end{figure}

\begin{figure}[t]
\centering
  \includegraphics[width=0.8\columnwidth]{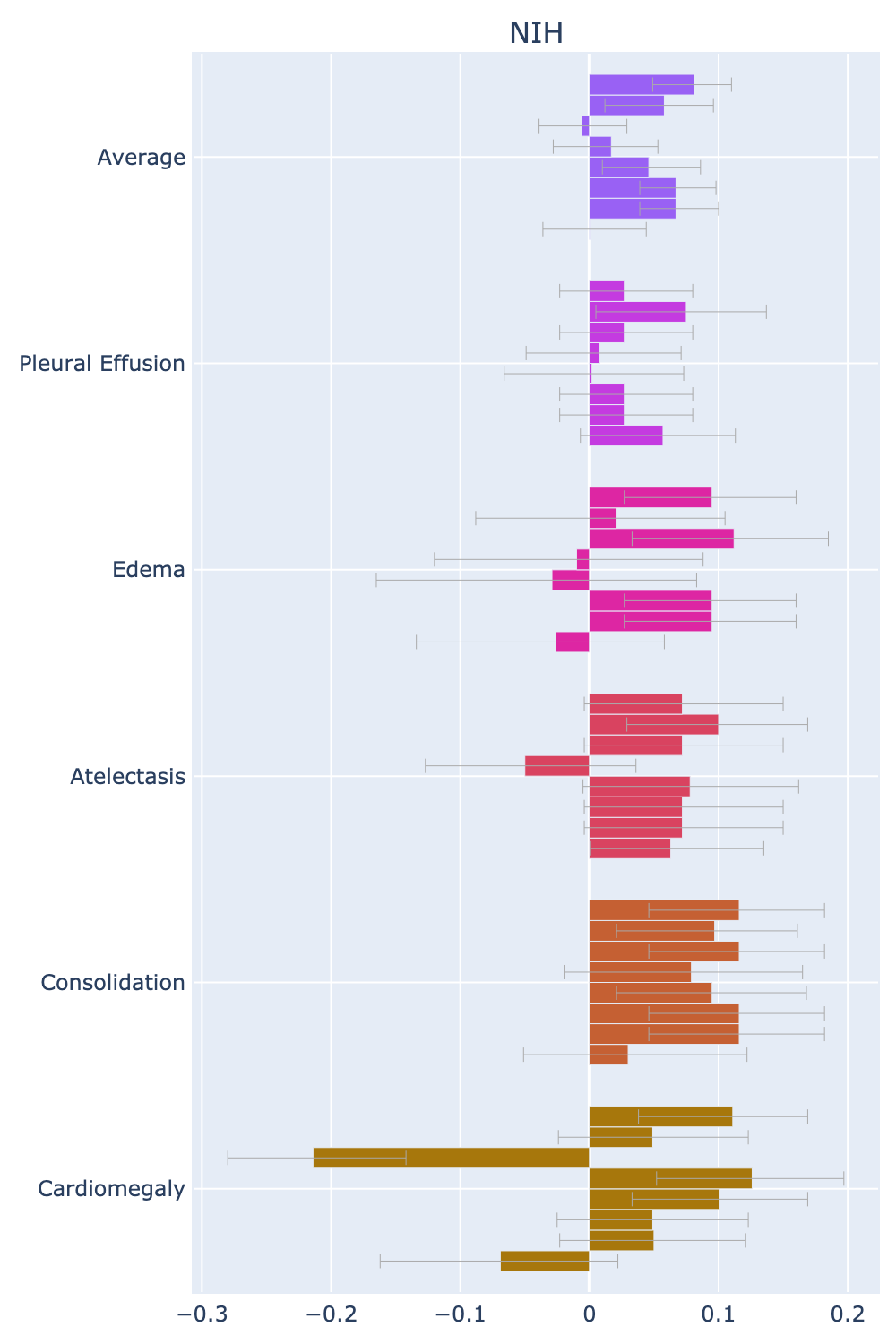}
  \caption{MCC differences in performance of the same models compared to another set of radiologists across the same pathologies on an external institution’s (NIH) data.}
\label{fig:nih}
\end{figure}

\subsubsection{Implication}
While using photos of chest X-rays to input into chest X-ray algorithms could enable any physician with a smartphone to get instant AI algorithm assistance, the performance of chest X-ray algorithms on photos of chest X-rays has not been thoroughly investigated. Several studies have highlighted the importance of generalizability of computer vision models with noise in  \citep{hendrycks_benchmarking_2019}. \citet{dodge_study_2017} demonstrated that deep neural networks perform poorly compared to humans on image classification on distorted images. \citet{geirhos_imagenet-trained_2019}, \citet{schmidt_adversarially_2018} have found that convolutional neural networks trained on specific image corruptions did not generalize, and the error patterns of network and human predictions were not similar on noisy and elastically deformed images.

\begin{table*}[t]
    \resizebox{\textwidth}{!}{  
\begin{tabular}{ll|r|rrrrr}
\toprule
Institution & Comparison            & Average              & Pleural   Effusion    & Edema                & Atelectasis          & Consolidation         & Cardiomegaly         \\
\midrule
CheXpert    & Radiologists          & 0.568 (0.542,0.597)  & 0.671 (0.618,0.727)   & 0.507 (0.431,0.570)  & 0.548 (0.496,0.606)  & 0.359 (0.262,0.444)   & 0.566 (0.511,0.620)  \\
    & Models                & 0.570 (0.543,0.599)  & 0.621 (0.575,0.670)   & 0.550 (0.474,0.637)  & 0.587 (0.529,0.640)  & 0.336 (0.264,0.418)   & 0.643 (0.584,0.695)  \\
    & Models - Radiologists & 0.002 (-0.028,0.030) & -0.05 (-0.092,-0.007) & 0.043 (-0.033,0.114) & 0.039 (-0.029,0.106) & -0.022 (-0.104,0.076) & 0.077 (0.040,0.135)  \\
\midrule
NIH         & Radiologists          & 0.537 (0.515,0.555)  & 0.642 (0.590,0.690)   & 0.618 (0.549,0.669)  & 0.469 (0.423,0.515)  & 0.455 (0.385,0.509)   & 0.492 (0.443,0.530)  \\
         & Models                & 0.578 (0.551,0.601)  & 0.673 (0.605,0.734)   & 0.662 (0.582,0.742)  & 0.529 (0.454,0.595)  & 0.551 (0.499,0.623)   & 0.517 (0.466,0.567)  \\
         & Models - Radiologists & 0.041 (0.010,0.072)  & 0.032 (-0.019,0.078)  & 0.044 (-0.028,0.124) & 0.060 (-0.003,0.126) & 0.096 (0.027,0.155)   & 0.025 (-0.028,0.078)\\
\bottomrule
\end{tabular}
 }
  \caption{MCC performance of models and radiologists on the CheXpert and NIH sets of chest X-rays, and their difference, with 95\% confidence intervals.}
\label{tab:tab3}
\end{table*}

\begin{figure*}[t]
\centering
  \includegraphics[width=\textwidth]{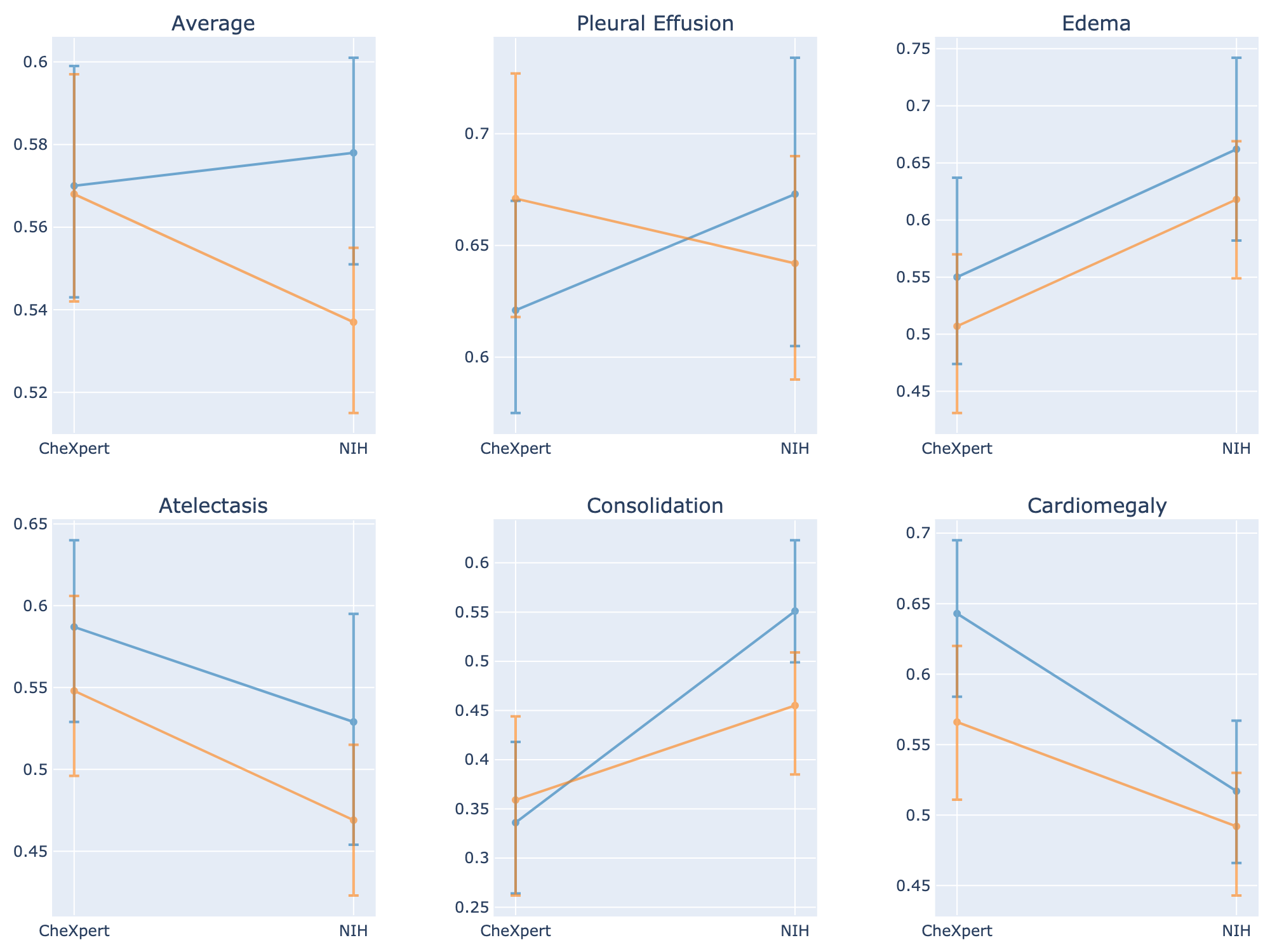}
  \caption{Overall change in performance of models (blue) and radiologists (orange) across CheXpert and the external institution dataset (NIH).}
\label{fig:overall}
\end{figure*}

\subsection{Comparison of Models and Radiologists on External Institution}

We measured the change in diagnostic performance of the same eight chest X-ray models on chest X-rays obtained at a different institution. We applied these models, trained on the CheXpert dataset from the Stanford Hospital, to a set of 420 frontal chest X-rays labeled as part of \citet{rajpurkar_deep_2018}. These X-rays are sourced from the ChestXray-14 dataset collected at the National Institutes of Health Clinical Center \citep{wang_chestX-ray8_2017}, and sampled to contain at least 50 cases of each pathology according to the original labels provided in the dataset. The reference standard on this set (NIH) was determined using a majority vote of three cardiothoracic subspecialty radiologists; six board-certified radiologists were used for comparison against the models. 

\subsubsection{Performance on external institution in comparison to radiologists}
On the external set (NIH), none of the models performed statistically significantly worse than radiologists (see Figure \ref{fig:nih}). On average over the pathologies, five models performed significantly better than radiologists. On specific pathologies, there were some models that achieved significantly better performance than radiologists: six models on consolidation, three models on cardiomegaly, four on edema, and two on atelectasis, one on pleural effusion.

\subsubsection{Implication}
Our finding that these models perform comparably to or at a level exceeding radiologists differs from a previous study which reported that a chest X-ray model failed to generalize to new populations or institutions separate from the training data, relying on institution specific and/or confounding cues to infer the label of interest \citep{zech_variable_2018}. Our findings may be attributed to the improvement in the generalizability of chest X-ray models owing to larger and higher-quality datasets that have been publicly released \citep{irvin_chexpert_2019, johnson_mimic-cxr_2019} Future work should investigate specific aspects of model training and dataset quality and size that lend to these differences, and whether self-supervised training procedures \citep{sowrirajan2020moco} increase generalizability across institutions.

\subsubsection{Performance change in context of radiologist performance}
Comparing performances on the CheXpert and NIH test sets, we found that on the NIH data set, in 16 instances models had a significantly better performance than radiologists; on the internal CheXpert test set, we observed that in 6 instances, models had a significantly higher performance than radiologists (see Figure \ref{fig:chexpert}). This difference may be attributed to a variety of factors including the difference in prevalence of pathologies or the difficulty in identifying them in the external test set compared to the internal set. We are able to contextualize the generalization ability of models to external institutions by comparing their differences to a radiologist performance benchmark, rather than provide a comparison of their absolute performances, which would not control for these possible differences. For instance, when considering cardiomegaly (see Figure \ref{fig:overall}), we observe a drop in model performance, which in isolation would indicate poor generalizability. However, in light of a similar drop in radiologist performance, we may be able to attribute the difference to differences in difficulties between the two datasets.

\section{Discussion}
The purpose of this work was to systematically address the key translation challenges for chest X-ray models in clinical application to common real-world scenarios. We found that several chest X-ray models had a drop in performance when applied to smartphone photos of chest X-rays, but even with this drop, some models still performed comparably to radiologists.  We also found that when models were tested on an external institution's data, they performed comparably to radiologists. In both forms of clinical distribution shifts we found that high-performance chest X-ray interpretation models trained on CheXpert produced clinically useful diagnostic performance.

Our work makes significant contributions over another investigation of chest X-ray models \citep{rajpurkar2020chexpedition}. While their study considered the differences in AUC of models when applied to photos of X-rays, they did not (1) compare the resulting performances against radiologists, (2) investigate the drop in performances on specific tasks, or (3) analyze drops in performances of individual models across tasks. Finally, while they compared the performance of models to radiologists on an external dataset, they did not investigate the change in performance of models between the internal dataset and the external dataset.

Strengths of our study include our systematic investigation of generalization performance of several chest X-ray models developed by different teams. Limitations of our work include that our study is still retrospective in nature, and prospective studies would further advance understanding of generalization under distribution shifts. Our systematic examination of the generalization capabilities of existing models can be extended to other tasks in medical AI \citep{duan_clinical_2019, huang2020penet, kiani_impact_2020, varma_automated_2019, topol_high-performance_2019}, and provide a framework for tracking technical readiness towards clinical translation. 

\bibliographystyle{ACM-Reference-Format}
\bibliography{sample-base}

\appendix

\end{document}